\documentclass[12pt]{iopart} 
\usepackage[pdftex]{graphicx}
\usepackage{iopams}
\usepackage{epstopdf}
\usepackage{graphicx}
\usepackage{cite}
\usepackage{gensymb}
\usepackage{tikz}
\usepackage{caption}
\usetikzlibrary{3d}

\begin{document}

\title{(Four) Dual Plaquette 3D Ising Models}

\author{Desmond A.  Johnston}
\address{School of Mathematical and Computer Sciences, Heriot-Watt University,
Riccarton, Edinburgh, EH14 4AS, Scotland}

\author{R. P. K. C. M. Ranasinghe}
\address{Department of Mathematics, University of Sri Jayewardenepura,
Gangodawila, Nugegoda 10250, Sri Lanka.}

\begin{abstract}
{A characteristic feature of the  $3d$ plaquette Ising model  is its planar subsystem symmetry. 
 The quantum version of this model has  been shown to be related via a duality to the X-Cube model, which has been paradigmatic in the new and rapidly developing field of fractons. The relation between the $3d$ plaquette Ising and the X-Cube model is similar to that between the   $2d$ quantum transverse spin Ising model  and the Toric Code. Gauging the {\it global} symmetry in the case of the $2d$ Ising model and considering the gauge invariant sector of the high temperature phase leads to the Toric Code, whereas gauging the {\it subsystem} symmetry of the $3d$ quantum transverse spin  plaquette Ising model  leads to the X-Cube model.  A non-standard dual formulation of the  $3d$ plaquette Ising model which utilises three flavours of spins has  recently been  discussed in the context of dualising the fracton-free sector of the X-Cube model.
In this paper we investigate the classical spin version of this non-standard dual Hamiltonian  and discuss its properties in relation to the more familiar Ashkin--Teller-like dual and further related dual formulations involving both  link and vertex spins and non-Ising spins.}
\end{abstract}

\def\be{\begin{equation}}
\def\ee{\end{equation}}
\def\bea{\begin{eqnarray}}
\def\eea{\end{eqnarray}}
\def\rp{r_{+}}
\def\rmm{r_{-}}

\newcommand{\Ham}{\mathcal{H}}     

\section{Introduction}

The Kramers--Wannier dual \cite{KW}  of the classical Ising Hamiltonian with nearest neighbour $\langle ij \rangle$ couplings
on a $3d$ cubic lattice
\begin{equation}
\label{e0I}
H_{Ising} =  -  \sum_{\langle ij \rangle}\sigma_{i} \sigma_{j}
\end{equation}
is the $3d$  Ising gauge theory
\begin{equation}
\label{e0G}
H_{Gauge} =  -  \sum_{\Box} U U U U 
\end{equation}
where the sum runs over plaquettes $\Box$ and the gauge spins $U$ live on the edges of the plaquettes. The coupling $\beta$ in the partition function $Z(\beta) = \sum_{\{ \sigma \}} \exp ( - \beta H_{Ising})$
and its dual $\beta^*$ 
in $Z(\beta^*) = \sum_{\{ U \}} \exp ( - \beta^* H_{Gauge})$
are related by $\beta^* = - \frac{1}{2} \log [ \tanh ( \beta  )]$. We use un-superscripted variables, e.g., $U, \sigma_i, \tau_i, \mu_i$,  to denote spins in classical Hamiltonians and superscripted variables, e.g., $\sigma^{x,z}_i, \tau^{x,z}_i, \mu^{x,z}_i$, to denote the Pauli matrices appearing in quantum Hamiltonians. The positional subscript indices $i,j,k\ldots$ are occasionally omitted for brevity.

In this paper we will investigate the relation between four (apparently) different
formulations of the dual to the $3d$ plaquette Ising model, which has also been dubbed the {\it gonihedric} Ising model  \cite{1a,1b,1c,1d} 
\begin{equation}
\label{e2k}
H_{\kappa=0} =  -  \sum_{\Box}\sigma_{i} \sigma_{j}\sigma_{k} \sigma_{l} \; .
\end{equation}
This,  like the $3d$ Ising gauge theory, has a plaquette $\Box$ interaction but  the spins now reside
at the vertices of a  $3d$ cubic lattice rather than on its edges. The subscript $\kappa=0$ appears because this plaquette
Hamiltonian is a particular case of a one-parameter family of gonihedric Hamiltonians
\begin{equation}
\label{e1}
H_{gonihedric} = - 4 \kappa \sum_{\langle ij\rangle }\sigma_{i} \sigma_{j}  +
\kappa \sum_{\langle \langle ij\rangle \rangle }\sigma_{i} \sigma_{j} 
- ( 1-\kappa )\sum_{\Box}\sigma_{i} \sigma_{j}\sigma_{k} \sigma_{l} \; .
\end{equation}
defined by Savvidy and Wegner \cite{2a,2b,2c,2d,2e,2f,2g,2h,2i,2j,2k}, where the $\langle \langle ij\rangle \rangle$
are next-to-nearest neighbour spin interactions.
The weights of spin cluster boundaries in this Hamiltonian are tuned to mimic
a gas of worldsheets arising from a gonihedric string action. 
When the gonihedric string worldsheets are discretized using triangulations, their action may be written as 
\begin{equation}
S = {1 \over 2} \sum_{\langle ij \rangle} | \vec X_i - \vec X_j | \; \theta (\alpha_{ij}),
\label{steiner}
\end{equation}
where
$\theta(\alpha_{ij}) = | \pi - \alpha_{ij} |$, $\alpha_{ij}$ is the dihedral angle between the
neighbouring triangles with a common edge $\langle ij \rangle$
and  $| \vec X_i - \vec X_j |$ are the lengths of the triangle edges. The $ \vec X_i$ give the embeddings of the vertices $i$ of the worldsheet discretization in the ambient spacetime.

The word gonihedric was originally coined to reflect the properties of this action,
which weights edge lengths  between non-coplanar triangles rather than the triangle areas, which is the case with a discretization of the standard Nambu--Goto/Polyakov string action.
It combines the Greek words gonia
for angle, referring to the dihedral angle, and hedra for base or face, referring
to the adjacent triangles. $H_{gonihedric}$ is an appropriate cubic lattice discretization 
for such an action because it too assigns zero weight to the areas of spin cluster
boundaries, weighting only edges and intersections \cite{3}. This gives $H_{gonihedric}$ very different properties
to $H_{Ising}$ where only the areas of spin cluster boundaries are weighted.

\begin{figure}[h!]
\centering
\begin{tikzpicture}[x  = {(0.5cm,0.5cm)},
                    y  = {(0.95cm,-0.25cm)},
                    z  = {(0cm,0.9cm)}]
\begin{scope}[canvas is yz plane at x=-1]
  \shade[left color=gray!70,right color=gray!20] (-1,-1) rectangle (1,1);
\end{scope}
\begin{scope}[canvas is xz plane at y=1]
  \shade[right color=gray!80,left color=gray!30] (-1,-1) rectangle (1,1);
\end{scope}
\begin{scope}[canvas is yx plane at z=1]
  \shade[top color=gray!70,bottom color=gray!20] (-1,-1) rectangle (1,1);
 \end{scope}
\end{tikzpicture}
\begin{tikzpicture}[x  = {(0.5cm,0.5cm)},
                    y  = {(0.95cm,-0.25cm)},
                    z  = {(0cm,0.9cm)}]
\begin{scope}[canvas is yz plane at x=-1]
  \shade[left color=gray!70,right color=gray!20] (-1,-1) rectangle (1,1);
\end{scope}
\begin{scope}[canvas is xz plane at y=1]
  \shade[right color=gray!80,left color=gray!30] (-1,-1) rectangle (1,1);
\end{scope}
\begin{scope}[canvas is yx plane at z=1]
  \shade[top color=black!100,bottom color=black!60] (-1,-1) rectangle (1,1);
\end{scope}
\end{tikzpicture}
\begin{tikzpicture}[x  = {(0.5cm,0.5cm)},
                    y  = {(0.95cm,-0.25cm)},
                    z  = {(0cm,0.9cm)}]
\begin{scope}[canvas is yz plane at x=-1]
  \shade[left color=black!100,right color=black!60] (-1,-1) rectangle (1,1);
\end{scope}
\begin{scope}[canvas is xz plane at y=1]
  \shade[right color=gray!70,left color=gray!20] (-1,-1) rectangle (1,1);
\end{scope}
\begin{scope}[canvas is yx plane at z=1]
  \shade[top color=gray!100,bottom color=gray!40] (-1,-1) rectangle (1,1);
\end{scope}
\end{tikzpicture}
\begin{tikzpicture}[x  = {(0.5cm,0.5cm)},
                    y  = {(0.95cm,-0.25cm)},
                    z  = {(0cm,0.9cm)}]
\begin{scope}[canvas is yz plane at x=-1]
  \shade[left color=gray!50,right color=gray!20] (-1,-1) rectangle (1,1);
\end{scope}
\begin{scope}[canvas is xz plane at y=1]
  \shade[right color=black!100,left color=black!60] (-1,-1) rectangle (1,1);
\end{scope}
\begin{scope}[canvas is yx plane at z=1]
  \shade[top color=gray!100,bottom color=gray!40] (-1,-1) rectangle (1,1);
\end{scope}
\end{tikzpicture}
\caption{Flipping the value of the Ising spins on a face of a single cube in 
 the $3d$ plaquette Ising Hamiltonian $H_{\kappa=0}$
 does not change its contribution to the energy. The first cube configuration is the ferromagnetic state with all spins $+$.  The  spins at the corners of the dark shaded faces on the other three are $-$, the others $+$.  All four of the
single cube  configurations shown have the same energy. } 
\label{fig:sketch}
\end{figure}
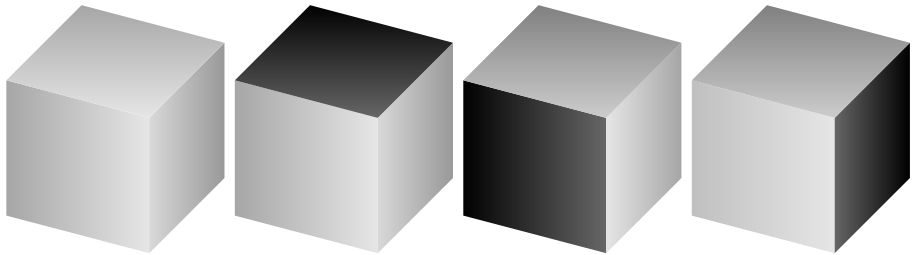

The $3d$ plaquette Ising action $H_{\kappa=0}$  has been shown to possess 
an exponentially (but sub-extensively)  degenerate low-temperature phase and a first order phase transition as well as interesting,
possibly glassy,  dynamical properties \cite{4a,4b,4c,4d,4e,4f,4g}. A characteristic feature is that  it displays a planar subsystem symmetry in which planes of spins may be flipped at zero energy cost, accounting for the degeneracy of the low temperature phase. This can be seen by looking at single cubes with a flipped face 
as in Figure \ref{fig:sketch} and  using these to tile the lattice. Since multiple faces can be flipped on the cube, intersecting planes of flipped spins are also possible.
The degeneracy affects the finite size scaling behaviour at the first order transition \cite{1sta,1stb,1stc,1std,1ste}, changing the universal $1/L^3$ finite size scaling shift in estimates of a first order  transition point on an $L^3$ lattice (with periodic boundary conditions) \cite{borgs-jankea,borgs-jankeb}  to $1/L^2$.
For non-zero $\kappa$ the planar subsystem symmetry appears to be broken at finite temperature \cite{pietig_wegnera,pietig_wegnerb}
and the transition becomes second order. The Kramers--Wannier dual to $H_{\kappa=0}$ takes the  form of an anisotropic Ashkin--Teller model  \cite{6}. It still possesses a planar subsystem symmetry and degenerate low temperature phase, so the modified finite size scaling at the first order transition is observed there also \cite{1sta,1stb,1stc,1std,1ste}.

The subsystem symmetry in the quantum spin version of the $3d$ plaquette Ising model has recently been shown to be closely linked to the properties of the X-Cube model \cite{VHF}, which has become a paradigmatic model for the new and rapidly developing field of fractons, which are quasiparticles with restricted mobility in isolation.  Some recent reviews of what is now a burgeoning fracton  literature can be found in \cite{fracton_review1,fracton_review2}.
To see the role played by the subsystem symmetry in constructing the X-Cube model, first consider  gauging the 
{\it global}  $\mathbb{Z}_2$ symmetry in the case of the $2d$  quantum transverse spin  Ising model 
\begin{equation}
H = - \beta \sum_{\langle ij \rangle} \sigma_i^z \sigma_j^z - h \sum_i \sigma_i^x \; .
\end{equation}
This can be done by introducing $\tau^z$  on the links and an additional plaquette flux term to endow the link spins with dynamics, which gives a gauge-invariant Ising
(or $\mathbb{Z}_2$ gauge--Higgs \cite{GH1,GH2,GH3}) model
 \begin{eqnarray}
 H & = & - \beta \sum_{\langle ij \rangle} \sigma_i^z  \tau^z \sigma_j^z - h \sum_i \sigma_i^x -   \beta_p \sum_{\Box} \tau^z \tau^z \tau^z \tau^z 
 \label{GI}
 \end{eqnarray}
 where we have dropped the link indices on the $\tau^z$  for conciseness.
The gauge-invariant sector  of  the high temperature phase, $\beta \to 0$, of  this model, where $ \sigma_i^x  \prod_{k \in +,i} \tau_k^x =1$ and  $k$ labels the four edges ($+$) incident to vertex $i$,
gives  Kitaev's Toric Code model \cite{TC_review,TC} 
\begin{eqnarray}
 H  & = &  -  h  \sum_i A_i  - \beta_p  \sum_{\Box} B_{\Box} \; .
  \end{eqnarray}
We use the gauge invariance to trade $\sigma_i^x$ for $\prod_{k \in +,i} \tau_k^x$, leaving  the mutually commuting terms
\begin{equation}
A_i = \prod_{k \in +,i} \tau_k^x,    \;  \; \; \;  B_{\Box} = \prod_{i \in \Box} \tau_i^z \; .
\end{equation}
and customarily set $h=\beta_p=1$. The Toric code displays topological order and has anyonic quasiparticle excitations.

\begin{figure}[h!]
\centering
\includegraphics[width=4cm]{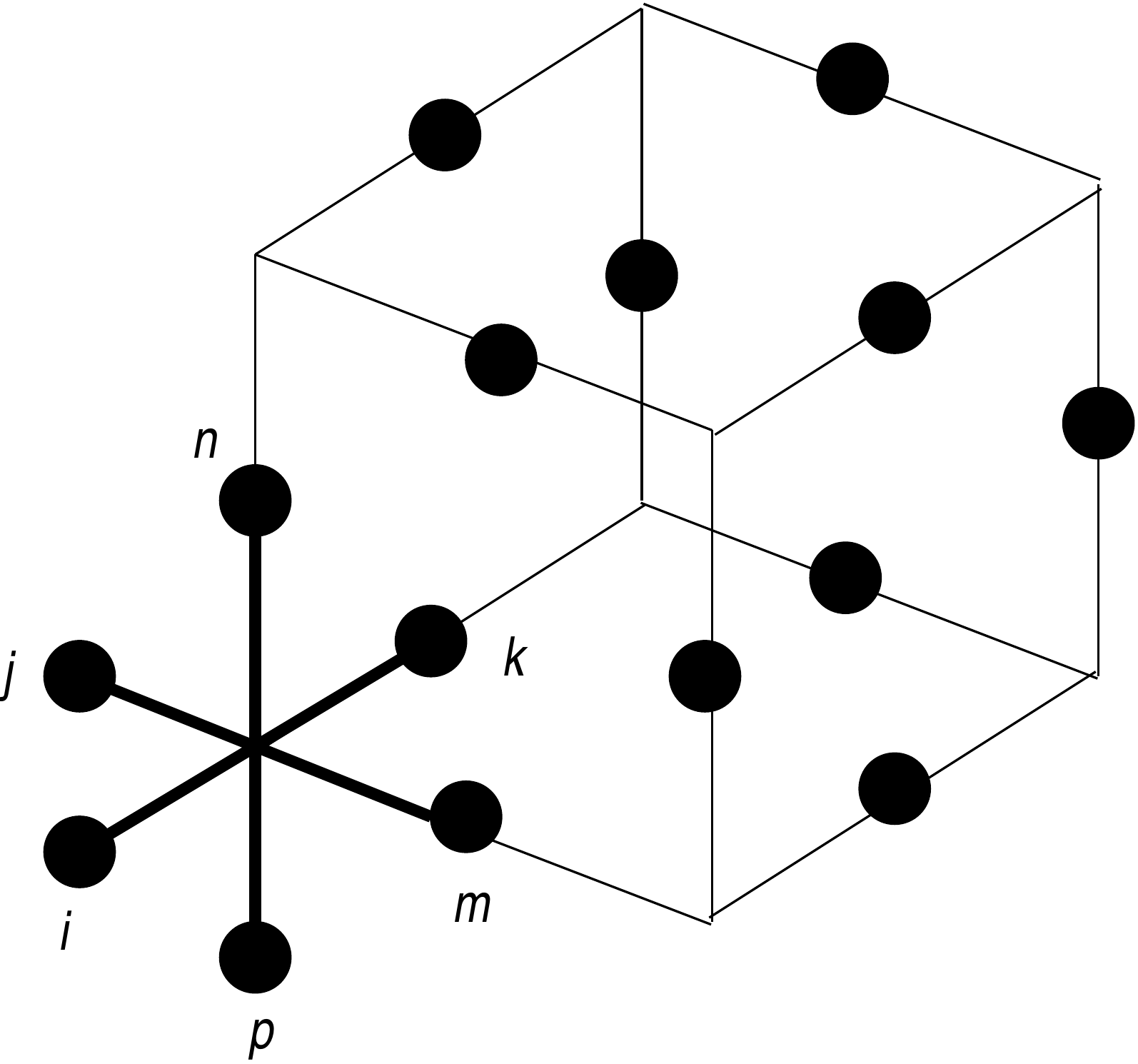}
\caption{The terms contributing to the X-Cube Hamiltonian. The cube $A$ term is a product of the twelve $\tau^x$ spins on the edges of the cube  and the three $B$   ``X'' terms  composed of $\tau^z$ spins lie in each of the three lattice planes as shown on the corner.}
\label{XcubeH}
\end{figure}

On the other hand, gauging the {\it subsystem} symmetry of the $3d$  plaquette Ising model  in a similar manner leads to the X-Cube model \cite{VHF}. In this case, when we start  with 
the quantum transverse spin $3d$ plaquette Ising model Hamiltonian
\begin{equation}
 H  = - \beta \sum_{\Box} \sigma_i^z \sigma_j^z  \sigma_k^z \sigma_l^z    -  h \sum_i \sigma_i^x
 \end{equation}
gauging the  $\mathbb{Z}_2$  {subsystem} symmetry requires inserting a $\tau^z$  which lives on the plaquettes
 \begin{eqnarray}
 H  & = & - \beta \sum_{\Box} \tau^z \sigma_i^z \sigma_j^z  \sigma_k^z \sigma_l^z  - h \sum_i \sigma_i^x   \; +  \;  \ldots  
 \end{eqnarray}
 The equivalent of the plaquette flux term in the Toric Code derivation is now a set of three ``X''  terms as shown in Figure~\ref{XcubeH}, one in each lattice plane
  $B^{xy,yz,xz}_{i} = \prod_{j \in +,i} \tau_j^z$.
 If we again consider the gauge invariant sector $\; \; \sigma_i^x  \prod_{k \in \Box,i } \tau_k^x =1$, where the $\tau_k^x$ live  on the twelve incident plaquettes impacted by flipping the single central spin at site $i$,   the high temperature limit
$\beta \to 0$ produces the X-Cube Hamiltonian
 \begin{eqnarray*}
  H  = -    \sum   A   -   \sum_i B_i^{xy}  -  \sum_i B_i^{yz}    -   \sum_i B_i^{xz}
\end{eqnarray*}
where it is simpler to think of the $\tau^x, \tau^z$'s  residing on the links of the  dual lattice. The $A$ term is a product of all the $\tau^x$  around a cube and the $B$ terms are the three ``crosses''  of $\tau^z$ 's shown
 in Figure~\ref{XcubeH}.
The remaining couplings have again been set to one.
The quasiparticles arising from defects in the $A$ terms are fractons and  cannot move in isolation, whereas the defects in the $B$ terms give lineons, which can only move in straight lines. The order in the X-Cube model is not topological. It has an exponential, but sub-extensive, ground state degeneracy  inherited from the plaquette Ising model as a consequence of the subsystem symmetry. 

It was observed recently  in \cite{odd}  that  the Hamiltonian for the fracton-free subsector (where all the $A$  cube terms are $+1$)  of the X-Cube model in a transverse field  
\begin{equation}
H = -   \sum_i B_i^{xy}  -  \sum_i B_i^{yz}    -   \sum_i B_i^{xz}  - g' \sum \tau^x
\end{equation} 
could be written in terms of a dual Hamiltonian
({at the risk of causing confusion we have kept the notation of} \cite{VHF} for the $A$ and $B$ terms rather than \cite{odd}, which swaps $A$ and $B$, though we denote the edge Pauli matrices by $\tau$ rather than $\sigma$ in distinction to both \cite{VHF,odd}) 
with three flavours of Ising spins $\sigma_i, \tau_i, \mu_i$ living on the vertices  of the cubic lattice rather than the links
\begin{eqnarray}
\label{duall2Q}
H = &-& g'\sum_{\langle ij \rangle} \sigma_i^z \sigma_j^z \mu_i^z \mu_j^z
- g'\sum_{\langle ik \rangle } \tau_i^z \tau_k^z  \mu_i^z \mu_k^z
- g'\sum_{ \langle jk \rangle} \sigma_j^z \sigma_k^z \tau_j^z \tau_k^z  \\
&-&    \sum_i  \left( \sigma_i^x \mu_i^x +  \tau_i^x \mu_i^x + \sigma_i^x \tau_i^x \right) \; ,\nonumber 
\end{eqnarray}
where the nearest neighbour sums in the four spin terms  each run along one of
the  orthogonal axes, with $ij, ik$ and $jk$ representing the
$z$, $y$ and $x$ axes respectively.  The constraint on the $A$ terms is automatically resolved by these spins.

In this paper we discuss the properties of the classical spin version of this Hamiltonian,\begin{equation}
\label{dual2}
H_{dual2} = - \sum_{\langle ij \rangle} \sigma_i \sigma_j \mu_i \mu_j
- \sum_{\langle ik \rangle } \tau_i \tau_k  \mu_i \mu_k
- \sum_{ \langle jk \rangle} \sigma_j \sigma_k \tau_j \tau_k \, ,
\end{equation}
 dubbed
$H_{dual2}$ for reasons to be explained in the next section. We shall see that it is closely related via a gauge-fixing to the Ashkin--Teller-like \cite{AT} Hamiltonian, $H_{dual1}$, constructed  using the classical Kramers--Wannier duality from the $3d$ plaquette Ising model, as well as though a decoration transformation to a  third Hamiltonian, $H_{dual3}$, which mixes edge and vertex spins. We find that the characteristic planar  subsystem symmetry of the $3d$ plaquette Ising model is still present in $H_{dual1,2,3}$ and also that the interesting, possibly glassy, dynamical properties of the $3d$ plaquette model are also apparent in the duals.  The Hamiltonians $H_{dual2,3}$ are already implicit in the discussion by Savvidy and Wegner in \cite{6b}
in the context of the general framework for dualities \cite{6c} in spin models.

\section{Duals Galore} 

\begin{figure}[h!]
\centering
\includegraphics[height=3cm]{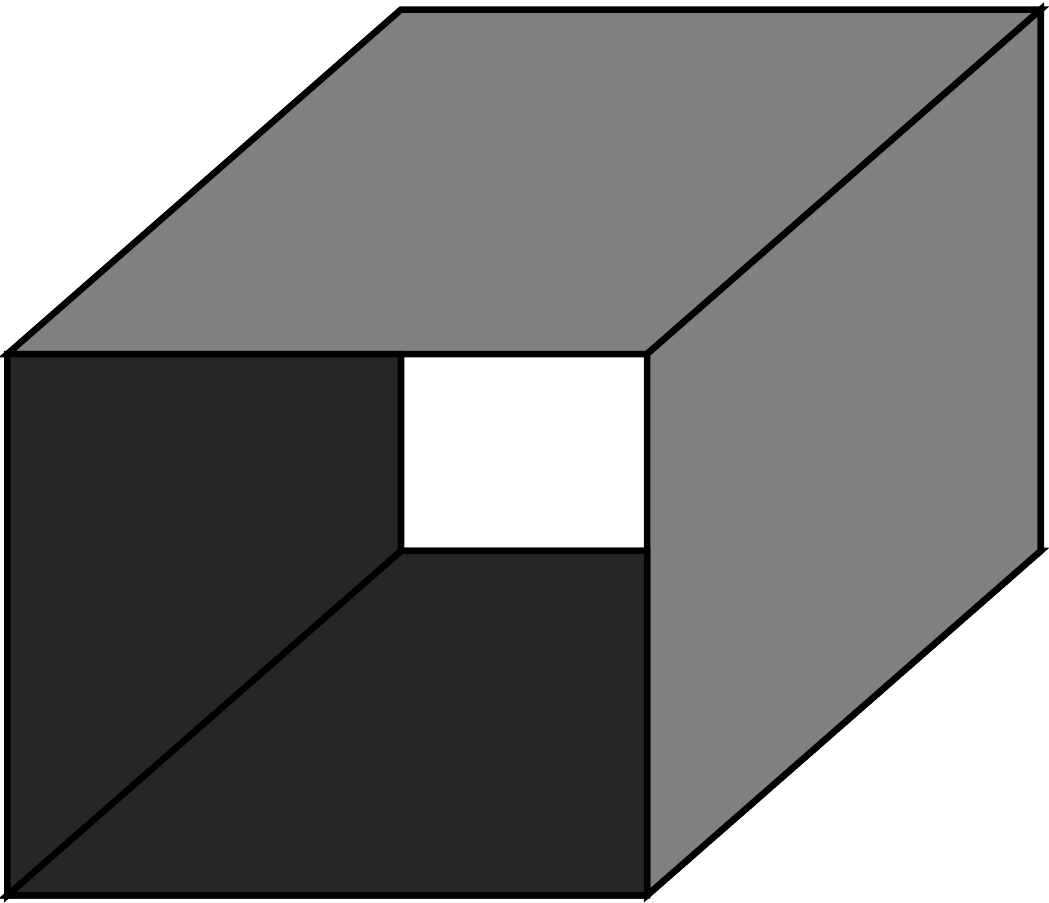}
\caption{One of the  matchbox surfaces which satisfy the algebra of Equation~(\ref{4thO}).}
\label{matchbox} 
\end{figure}

The Kramers--Wannier dual to $H_{\kappa=0}$ was initially constructed  by Savvidy et al. \cite{6} 
by considering the high temperature expansion of the plaquette Hamiltonian 
\begin{eqnarray}
Z (\beta)  &=& \sum_{\{\sigma\}}  \exp (- \beta H_{\kappa=0} ) \nonumber \\
&=& \sum_{\{\sigma\}} \prod_{\Box} \cosh  \left(\beta \right) \left[1 + \tanh \left(\beta \right) ( \sigma_i \sigma_j \sigma_k \sigma_l )\right] 
\label{z2}
\end{eqnarray}
which can be written as
\begin{equation}
Z (\beta)  = \left[ 2 \cosh \left(\beta \right) \right]^{3 L^3} \sum_{\{ S \}} \left[\tanh \left(\beta \right)\right]^{n ( S ) } 
\label{z2a}
\end{equation}
on an $L^3$ cubic lattice, where the sum runs over closed surfaces with an even number of plaquettes at any vertex. In the summation $n ( S )$ is the number of 
plaquettes in a given surface. 
The low temperature expansion, i.e., high temperature in the dual variable 
$$\beta^* = - (1/2)  \log [  \tanh ( \beta ) ]$$
of the following
{\it anisotropic} Hamiltonian 
\begin{equation}
\label{dual0}
 H_{dual0} = - \sum_{\langle ij \rangle} \sigma_{i}  \sigma_{j} 
- \sum_{\langle ik \rangle } \tau_{i}  \tau_{k} 
-  \sum_{ \langle jk \rangle} \eta_{j} \eta_{k} 
\end{equation}
produced the requisite diagrams. In $H_{dual0} $ the sums are one-dimensional and run along
the  orthogonal axes, with $ij, ik$ and $jk$ again representing the
$z$, $y$ and $x$ axes respectively using our conventions. 
The spins are non-Ising and live in the fourth order Abelian group, since
the geometric constraints on having an even number of plaquettes at each vertex mean that
\begin{eqnarray}
\label{4thO}
e \sigma &=&  \sigma \; , \;  \; e \tau = \tau \; ,  \; \; e \eta = \eta \nonumber \\	
\sigma^2 &=& \tau^2 = \eta^2 = e \\
\sigma \tau &=& \eta \, , \; \; \tau \eta = \sigma \; , \; \; \eta \sigma = \tau \nonumber
\end{eqnarray}
with $e$ being the identity element. They can be thought of as representing differently oriented
matchbox surfaces such as that shown in Figure~\ref{matchbox}, which are combined by facewise multiplication. 
The shaded
faces carry a negative sign and the associated spin variable lives at the centre
of the matchbox. Any spin cluster boundary in the model can be constructed from such matchboxes while still satisfying the local constraint on the number of incident plaquettes.

The spins may also be taken to be Ising ($\pm 1$) variables
if we set $\eta_i = \sigma_i \; \tau_i $, which is more convenient for simulations.
This modifies $H_{dual0}$ to an anisotropically coupled Ashkin--Teller Hamiltonian \cite{AT}
\begin{equation} 
\label{dual1}
H_{dual1} = -  \sum_{ \langle ij \rangle} \sigma_{i}  \sigma_{j} 
-  \sum_{ \langle ik \rangle } \tau_{i}  \tau_{k} 
-   \sum_{\langle jk \rangle} \sigma_{j} \sigma_{k} \tau_{j}  \tau_{k} \, . 
\end{equation}
This formulation of the dual model was first investigated numerically in \cite{5} and it was found that it
displayed a first order phase transition and a similar planar subsystem symmetry to that of $H_{\kappa=0}$. The continued presence of the subsystem symmetry
was a consequence of the anisotropic couplings, which allowed a greater freedom in transforming the spin variables than in the isotropically coupled version of Equation~(\ref{dual1}), which is just the Ashkin--Teller model at its four-state Potts point.

It is  possible to construct $H_{dual1}$  and its higher dimensional equivalents \cite{6b} using the  general framework
for duality in Ising lattice spin models that was first
formulated by Wegner in \cite{6c}. Suprisingly, there are two further possible
ways to write the dual to $H_{\kappa=0}$ in three dimensions with this machinery, using either
the general formula for the dual of codimension one surfaces or
the  formula for the dual of two dimensional surfaces in $d$ dimensions. 
If we temporarily use the notation of \cite{6b},
the dual Hamiltonian
for a codimension one surface in $d$ dimensions is given there by 
\bea
\label{codim1}
H^{d}_{dual,  codim 1} &=& - \sum_{\alpha<\beta, \,  \vec r}\prod_{\gamma}
\Lambda_{\alpha,\beta\gamma}(\vec r)
\Lambda_{\alpha,\beta\gamma}(\vec r+\vec e_{\gamma})
\Lambda_{\beta,\alpha\gamma}(\vec r)
\Lambda_{\beta,\alpha\gamma}(\vec r+\vec e_{\gamma}) \nonumber \\
{}
\eea
where the $\Lambda$ spins live on each of the $(d-3)$ dimensional (hyper)vertices 
situated at the vertices $\vec r$ of the hypercubic lattice and the indices $\alpha, \beta,\gamma$
run from $1$ to $d$. The unit vectors  $\vec e_{\gamma}$ point along the lattice axes. 
On the other hand, the
dual Hamiltonian for a two-dimensional gonihedric surface embedded in $d$ dimensions is of the form 
\be
\label{d2d}
H^{d}_{dual, \, 2d} = -\sum_{\vec r}\sum_{\beta \neq \gamma}
\Lambda_{\beta\gamma}(\vec r)
\Gamma(\vec r,\vec r +\vec e_{\gamma})
\Lambda_{\beta\gamma}(\vec r+\vec e_\gamma)
\ee
where we now have $\Gamma$ spins on each (hyper)edge
in addition to the $\Lambda$ spins at each vertex.

If we  specialize to two dimensional plaquette surfaces embedded in a cubic lattice in three dimensions, which is the case
for the dual of $H_{\kappa=0}$, either formulation may be employed since this is both a codimension one surface and a two-dimensional surface 
embedded in three dimensions.
Returning to our own notation, the codimension one Hamiltonian of Equation~(\ref{codim1}) in three dimensions may be written as 
\begin{equation*}
\label{dual3ab}
H_{dual2} = - \sum_{\langle ij \rangle} \sigma_i \sigma_j \mu_i \mu_j
- \sum_{\langle ik \rangle } \tau_i \tau_k  \mu_i \mu_k
- \sum_{ \langle jk \rangle} \sigma_j \sigma_k \tau_j \tau_k \, ,
\end{equation*}
which is just the Hamiltonian of Equation~(\ref{dual2}) that appeared as the classical spin limit of the dual to the fracton-free
subspace of the X-Cube model.
The  three
flavours of spins living at each vertex  display a local Ising gauge symmetry
$
\sigma_i, \tau_i, \mu_i \to \gamma_i \sigma_i, \gamma_i \tau_i, \gamma_i \mu_i
$
in addition to the planar subsystem symmetry shared with $H_{\kappa=0}$ and  $H_{dual1}$, as we shall see presently. 

Still within the general approach of  Savvidy and Wegner \cite{6b,6c}, in three dimensions the Hamiltonian of Equation~(\ref{d2d}) for the two-dimensional surface variant   also contains three flavours of vertex spins
$\sigma_i, \tau_i , \mu_i$, but in addition there are gauge-like spin variables $U_{ij}^{1,2,3}$ living on the  lattice edges 
which couple in an anisotropic manner to the vertex spins 
\bea
\label{dual3}
H_{dual3} &=& - \sum_{\langle ij \rangle} \left( \sigma_i U^{1}_{ij} \sigma_j  + \mu_i U^{1}_{ij} \mu_j \right) 
- \sum_{\langle ik \rangle } \left( \tau_i U^{2}_{ik} \tau_k +  \mu_i  U^{2}_{ik} \mu_k \right) \nonumber \\
&-&  \sum_{ \langle jk \rangle} \left( \sigma_j U^{3}_{jk} \sigma_k  + \tau_j U^{3}_{jk} \tau_k  \right) \; .
\eea 
We thus have four different Hamiltonian formulations for the dual of the plaquette Hamiltonian $H_{\kappa=0}$ in three dimensions: 
\begin{itemize}
\item{} $H_{dual0}$ in Equation~(\ref{dual0}) with non-Ising spins.
\item{} $H_{dual1}$ in Equation~(\ref{dual1}) with Ising spins, which is Ashkin--Teller in form.	
\item{} $H_{dual2}$ in Equation~(\ref{dual2}) containing purely four spin interactions.
\item{} $H_{dual3}$ in Equation~(\ref{dual3}) containing both vertex spins  and gauge-like edge spins.
\end{itemize} 
We have already seen that setting $\eta_i = \sigma_i \; \tau_i $ in $H_{dual0}$ in  equ.~(\ref{dual0}), with $\sigma_i, \tau_i$ being Ising spins, keeps the algebra of
Equation~(\ref{4thO}) intact and gives the Ashkin--Teller  Hamiltonian of $H_{dual1}$ in Equation~(\ref{dual1}).
In the next section we discuss
the relation between the four-spin Hamiltonian $H_{dual2}$ of Equation~(\ref{dual2}) and the gauge-spin Hamiltonian $H_{dual3}$ of Equation~(\ref{dual3}), and thereafter that between $H_{dual2}$
and $H_{dual1}$.

\section{Decoration}

\begin{figure}[h!]
\centering
\includegraphics[height=3cm]{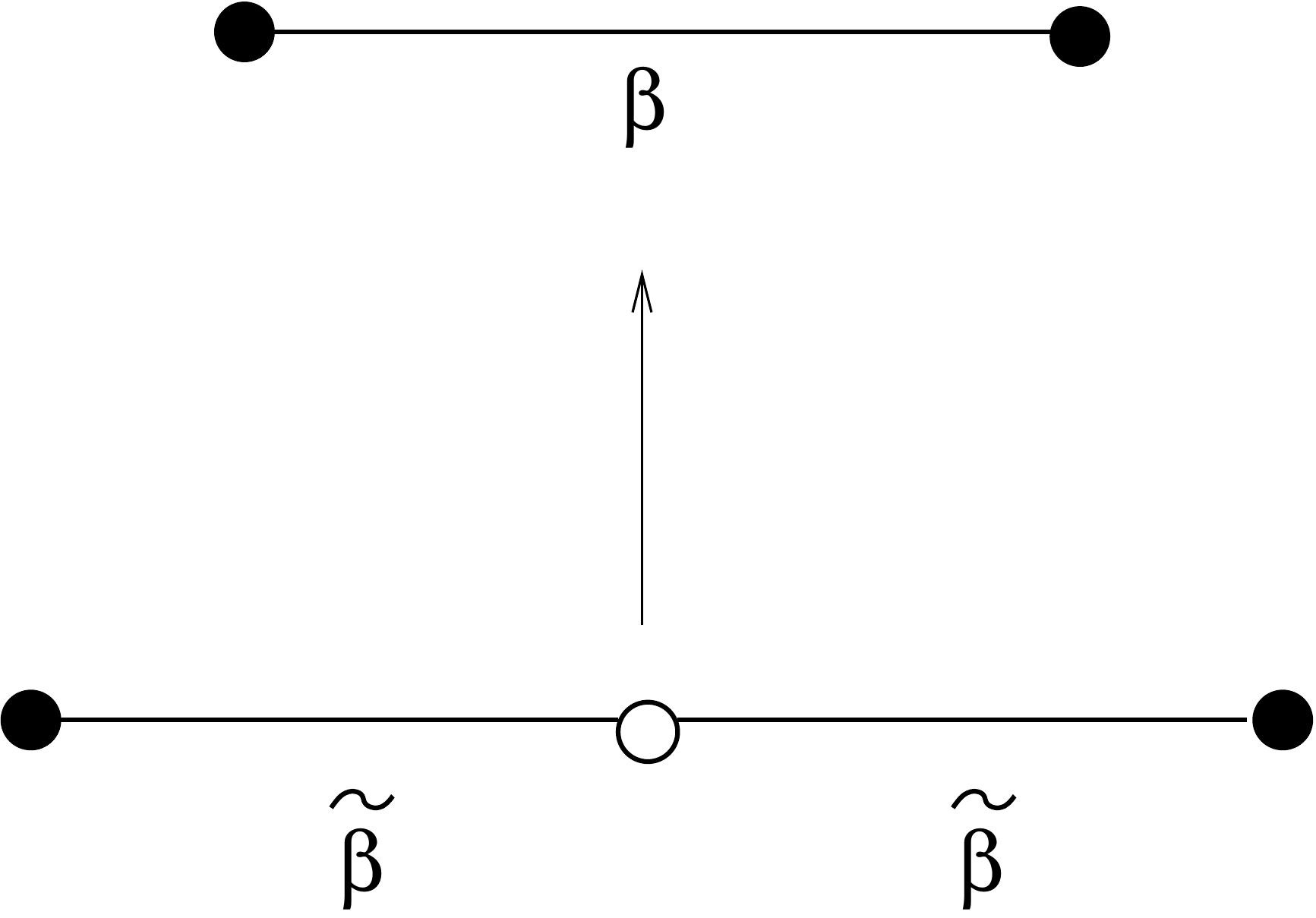}
\caption{The standard decoration transformation. Summing over the central spin $s$ denoted by an open dot  gives a new effective coupling
$\beta  =  \frac{1}{2}  \log [  \cosh ( 2 \tilde \beta )]$ between the spins on the end of the link.  }
\label{fig0a}
\end{figure}

The equivalence between $H_{dual3}$ and $H_{dual2}$  is a
consequence of a variation of the classical decoration transformation \cite{7}. In the standard transformation
an edge with spins $\sigma_1, \sigma_2$ at each vertex is decorated with a link spin $s$
as in Figure~\ref{fig0a}.
If the coupling between $s$ and $\sigma_{1}$ and $\sigma_{2}$ is $\tilde \beta$, 
summing over the central spin $s$ gives rise to a 
new effective coupling $\beta$ between the 
primary vertex spins $\sigma_1, \sigma_2$
\begin{equation}
\label{decor}
\sum_{s} \exp \left[ \tilde \beta s ( \sigma_1 + \sigma_2 ) \right] = A \, \exp ( \beta \sigma_1 \sigma_2).
\end{equation}
Both the prefactor $A$ and the coupling $\beta$ may be expressed in terms of $\tilde \beta$
by enumerating possible spin configurations in Equation~(\ref{decor}).
This gives 
\bea
A &=& 2  \cosh ( 2 \tilde \beta )^{1/2} \nonumber \\
\beta &=&   \frac{1}{2}  \log [  \cosh ( 2 \tilde \beta )] .
\label{e2}
\eea
We can repeat this procedure with the $U$ spins on each edge in $H_{dual3}$. In this case
each direction has two flavours of vertex spin and performing the sum generates
the four-spin couplings of $H_{dual2}$, for example
\bea
\sum_{\{U^1_{12}\}} \exp \left[ \tilde \beta   \left( \sigma_1 U^1_{12} \sigma_2  + \mu_1 U^1_{12} \mu_2 \right) \right]
= A  \, \exp ( \beta \sigma_1 \sigma_2 \mu_1  \mu_2) .
\eea
where $A$ and the relation between $\beta, \tilde \beta$ are the same as in Equation~(\ref{e2}).

The sum over $U$ may be carried out globally over every edge which immediately demonstrates
equivalence of the partition functions for $H_{dual3}$ and $H_{dual2}$
\bea
\label{sUs2}
Z &=& \sum_{\{U, \sigma\}} \exp [ -\tilde \beta H_{dual3} ] \nonumber \\
 &=& \sum_{\{U, \sigma\}} \exp \left [ \tilde \beta\sum_{\langle ij \rangle} \left( \sigma_i U^{1}_{ij} \sigma_j  + \mu_i U^{1}_{ij} \mu_j \right) 
+ \tilde \beta \sum_{\langle ik \rangle } \left( \tau_i U^{2}_{ik} \tau_k +  \mu_i  U^{2}_{ik} \mu_k \right) \nonumber \right. \\
&+& \left. \tilde \beta \sum_{ \langle jk \rangle} \left( \sigma_j U^{3}_{jk} \sigma_k  + \tau_j U^{3}_{jk} \tau_k  \right) \right] \\
&=& B \, \sum_{\{\sigma\}}  \exp \left[  \beta \left( \sum_{\langle ij \rangle} \sigma_i \sigma_j \mu_i \mu_j + \sum_{\langle ik \rangle } \tau_i \tau_k  \mu_i   \mu_k  +  \sum_{ \langle jk \rangle}  \sigma_j  \sigma_k   \tau_j \tau_k \right)  \right] \nonumber \\
&=& B \, \sum_{\{\sigma\}} \exp [ -\beta H_{dual2} ] \nonumber .
\eea 
The overall factor
$B$ coming from a product of $A$'s on the individual links
is irrelevant for calculating physical quantities and the two couplings are again related by the 
decoration relation, $\beta =   \frac{1}{2}  \log  [ \cosh ( 2 \tilde \beta ) ]$.

\section{Gauge Fixing and Subsystem Symmetry}

The equivalence between $H_{dual2}$ and $H_{dual1}$, on the other hand, is a consequence of a
gauge symmetry which is present in $H_{dual2}$  \cite{5b} 
\be
\sigma_i, \, \tau_i, \, \mu_i \, \to \,  \gamma_i  \sigma_i, \,  \gamma_i \tau_i, \, \gamma_i \mu_i \, .
\label{local_gauge}
\ee
We are at liberty to choose the Ising spin gauge transformation parameter $\gamma_i$ to be equal to 
one of the spin values, say $\mu_i$, at each site
so the gauge transformation then becomes
\be
\sigma_i, \, \tau_i, \, \mu_i \, \to \,  \mu_i  \sigma_i, \,  \mu_i \tau_i, \, 1
\ee
which, using the fact that the sum over the remaining spin variables $\sigma_i, \tau_i$ is invariant under the transformation, relates the partition functions for the two Hamiltonians as
\bea 
Z &=& \sum_{ \{\sigma,\tau,\mu \}} \exp \left[ - \beta H_{dual2} ( \sigma,\tau,\mu) \right] \nonumber \\
&=& 2^{L^3} \sum_{\{ \sigma,\tau \}} \exp \left[ - \beta H_{dual2} ( \sigma,\tau,\mu=1) \right] \\
&=&  2^{L^3} \sum_{\{\sigma,\tau \}} \exp \left[ - \beta H_{dual1} ( \sigma,\tau) \right] \; .
\nonumber
\eea
The coupling $\beta$ is not transformed in this case and we can, of course, choose to eliminate any one of the 
three spins, which simply amounts to relabelling the axes. From this perspective $H_{dual1}$ is simply a gauge-fixed
version of $H_{dual2}$. This can be  confirmed by Monte-Carlo simulations which measure the same energies (and energy distributions)
and transition points for the observed first order phase transitions \cite{5b}.

The equivalence between $H_{dual3}$ and $H_{dual2}$ described in the preceding section via the decoration transformation also sheds light on the presence of this gauge symmetry in  $H_{dual2}$. All the terms in $H_{dual3}$ are of the gauge-matter coupling form $ \sigma_i U_{ij} \sigma_j$, so this action possesses a similar, standard gauge invariance to that seen in other gauge-matter systems such as 
the  gauge--Ising model of Equation~(\ref{GI}), namely
\bea    
\sigma_i \to \gamma_i \sigma_i \; , \; \sigma_j \to \gamma_j \sigma_j \; , \; U_{ij}^{1,3} \to \gamma_i U_{ij}^{1,3} \gamma_j \nonumber \\
\tau_i \to \gamma_i \tau_i \; , \; \tau_j \to \gamma_j \tau_j \; , \; U_{ij}^{2,3} \to \gamma_i U_{ij}^{2,3} \gamma_j \\
\mu_i \to \gamma_i \mu_i \; , \; \mu_j \to \gamma_j \mu_j \; , \; U_{ij}^{1,2} \to \gamma_i U_{ij}^{1,2} \gamma_j \nonumber \; .
\eea
When the $U$ spins are summed over to give  $H_{dual2}$, the gauge symmetry of the $\sigma, \tau$ and $\mu$
spins in Equation~(\ref{local_gauge}) remains as an echo of this symmetry. In both cases if we look
at a single site transformation all three spins $\sigma_i, \tau_i$ and $\mu_i$ must be transformed. In  $H_{dual3}$ this is a consequence of the way in which the three edge spins $U^{1,2,3}_{ij}$ couple to the vertex spins.

\begin{figure}[h!]
\centering
\begin{tikzpicture}
        \draw [->] (0,0) -- (0,1.5);
        \node at (0,1.7) (nodesm) {$\sigma \, \mu$};
        \draw [->] (0,0) -- (1.4,0);
         \node at (1.7,0) (nodest) {$\sigma \, \tau$};
        \draw [->] (0,0) -- (1.0,0.5);
        \node at (1.2,0.7) (nodest) {$\tau \, \mu$};
    \end{tikzpicture}
\begin{tikzpicture}[x  = {(0.5cm,0.5cm)},
                    y  = {(0.95cm,-0.25cm)},
                    z  = {(0cm,0.9cm)}]
\begin{scope}[canvas is yz plane at x=-1]
  \shade[left color=gray!70,right color=gray!20] (-1,-1) rectangle (1,1);
\end{scope}
\begin{scope}[canvas is xz plane at y=1]
  \shade[right color=gray!80,left color=gray!30] (-1,-1) rectangle (1,1);
\end{scope}
\begin{scope}[canvas is yx plane at z=1]
  \shade[top color=gray!70,bottom color=gray!20] (-1,-1) rectangle (1,1);
 \end{scope}
\end{tikzpicture}
\begin{tikzpicture}[x  = {(0.5cm,0.5cm)},
                    y  = {(0.95cm,-0.25cm)},
                    z  = {(0cm,0.9cm)}]
\begin{scope}[canvas is yz plane at x=-1]
  \shade[left color=gray!70,right color=gray!20] (-1,-1) rectangle (1,1);
\end{scope}
\begin{scope}[canvas is xz plane at y=1]
  \shade[right color=gray!80,left color=gray!30] (-1,-1) rectangle (1,1);
\end{scope}
\begin{scope}[canvas is yx plane at z=1]
  \shade[top color=black!100,bottom color=black!60] (-1,-1) rectangle (1,1);
\end{scope}
\node[anchor=south] at (current bounding box.north) 
      {$+ - +$};
\end{tikzpicture}
\begin{tikzpicture}[x  = {(0.5cm,0.5cm)},
                    y  = {(0.95cm,-0.25cm)},
                    z  = {(0cm,0.9cm)}]
\begin{scope}[canvas is yz plane at x=-1]
  \shade[left color=black!100,right color=black!60] (-1,-1) rectangle (1,1);
\end{scope}
\begin{scope}[canvas is xz plane at y=1]
  \shade[right color=gray!70,left color=gray!20] (-1,-1) rectangle (1,1);
\end{scope}
\begin{scope}[canvas is yx plane at z=1]
  \shade[top color=gray!100,bottom color=gray!40] (-1,-1) rectangle (1,1);
\end{scope}
\node[anchor=south] at (current bounding box.north) 
      {$- + +$};
\end{tikzpicture}
\begin{tikzpicture}[x  = {(0.5cm,0.5cm)},
                    y  = {(0.95cm,-0.25cm)},
                    z  = {(0cm,0.9cm)}]
\begin{scope}[canvas is yz plane at x=-1]
  \shade[left color=gray!50,right color=gray!20] (-1,-1) rectangle (1,1);
\end{scope}
\begin{scope}[canvas is xz plane at y=1]
  \shade[right color=black!100,left color=black!60] (-1,-1) rectangle (1,1);
\end{scope}
\begin{scope}[canvas is yx plane at z=1]
  \shade[top color=gray!100,bottom color=gray!40] (-1,-1) rectangle (1,1);
\end{scope}
\node[anchor=south] at (current bounding box.north) 
      {$+  +  -$};
\end{tikzpicture}
\caption{Four Possible ground state spin configurations on a cube for $H_{dual2}$. The initial cube again has all $+$ spins and  the $\sigma,\tau,\mu$ values are shown for the spins at the corners of the darker shaded flipped faces, with the other spins being positive. The directions of the anisotropic couplings in the Hamiltonian are indicated. }
\label{groundstate_shade} 
\end{figure}
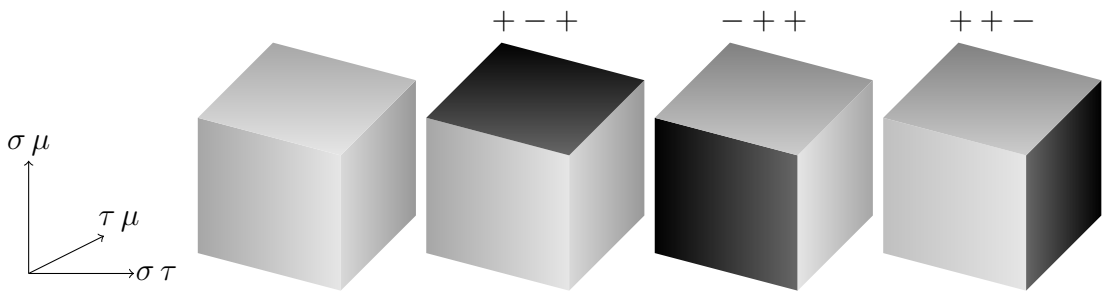

A characteristic feature of $H_{dual1}$ is the planar subsystem symmetry  intermediate
between a gauge and a global symmetry,  just as with the $3d$ plaquette Ising Hamiltonian. For  $H_{dual2}$   the anisotropic couplings mean that it is still possible to flip planes of one of the spins (the one which is ``missing'' from the interactions in the direction perpendicular to the planes)
at zero energy cost as shown in Figure~\ref{groundstate_shade}.
It is also possible to flip two or three orthogonal faces on the cube, so tiling the entire lattice with
such combinations we can see that in addition to the purely ferromagnetic ground state we  may have
arbitrary (and possibly intersecting) flipped planes  of spins. 

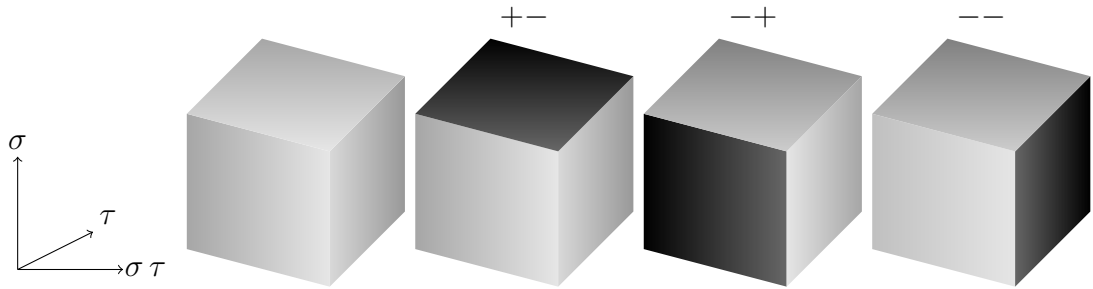
\begin{figure}[h!]
\centering
\begin{tikzpicture}
        \draw [->] (0,0) -- (0,1.5);
        \node at (0,1.7) (nodesm) {$\sigma$};
        \draw [->] (0,0) -- (1.4,0);
         \node at (1.7,0) (nodest) {$\sigma \, \tau$};
        \draw [->] (0,0) -- (1.0,0.5);
         \node at (1.2,0.7) (nodest) {$\tau$};
    \end{tikzpicture}
\begin{tikzpicture}[x  = {(0.5cm,0.5cm)},
                    y  = {(0.95cm,-0.25cm)},
                    z  = {(0cm,0.9cm)}]
\begin{scope}[canvas is yz plane at x=-1]
  \shade[left color=gray!70,right color=gray!20] (-1,-1) rectangle (1,1);
\end{scope}
\begin{scope}[canvas is xz plane at y=1]
  \shade[right color=gray!80,left color=gray!30] (-1,-1) rectangle (1,1);
\end{scope}
\begin{scope}[canvas is yx plane at z=1]
  \shade[top color=gray!70,bottom color=gray!20] (-1,-1) rectangle (1,1);
 \end{scope}
\end{tikzpicture}
\begin{tikzpicture}[x  = {(0.5cm,0.5cm)},
                    y  = {(0.95cm,-0.25cm)},
                    z  = {(0cm,0.9cm)}]
\begin{scope}[canvas is yz plane at x=-1]
  \shade[left color=gray!70,right color=gray!20] (-1,-1) rectangle (1,1);
\end{scope}
\begin{scope}[canvas is xz plane at y=1]
  \shade[right color=gray!80,left color=gray!30] (-1,-1) rectangle (1,1);
\end{scope}
\begin{scope}[canvas is yx plane at z=1]
  \shade[top color=black!100,bottom color=black!60] (-1,-1) rectangle (1,1);
\end{scope}
\node[anchor=south] at (current bounding box.north) 
      {$+ - $};
\end{tikzpicture}
\begin{tikzpicture}[x  = {(0.5cm,0.5cm)},
                    y  = {(0.95cm,-0.25cm)},
                    z  = {(0cm,0.9cm)}]
\begin{scope}[canvas is yz plane at x=-1]
  \shade[left color=black!100,right color=black!60] (-1,-1) rectangle (1,1);
\end{scope}
\begin{scope}[canvas is xz plane at y=1]
  \shade[right color=gray!70,left color=gray!20] (-1,-1) rectangle (1,1);
\end{scope}
\begin{scope}[canvas is yx plane at z=1]
  \shade[top color=gray!100,bottom color=gray!40] (-1,-1) rectangle (1,1);
\end{scope}
\node[anchor=south] at (current bounding box.north) 
      {$- + $};
\end{tikzpicture}
\begin{tikzpicture}[x  = {(0.5cm,0.5cm)},
                    y  = {(0.95cm,-0.25cm)},
                    z  = {(0cm,0.9cm)}]
\begin{scope}[canvas is yz plane at x=-1]
  \shade[left color=gray!50,right color=gray!20] (-1,-1) rectangle (1,1);
\end{scope}
\begin{scope}[canvas is xz plane at y=1]
  \shade[right color=black!100,left color=black!60] (-1,-1) rectangle (1,1);
\end{scope}
\begin{scope}[canvas is yx plane at z=1]
  \shade[top color=gray!100,bottom color=gray!40] (-1,-1) rectangle (1,1);
\end{scope}
\node[anchor=south] at (current bounding box.north) 
      {$-  -$};
\end{tikzpicture}
\caption{Four possible ground state spin configurations on a cube for the Ashkin--Teller formulation of the dual
Hamiltonian, $H_{dual1}$. The $\sigma,\tau$ values are shown for the darker shaded flipped planes. The directions of the anisotropic couplings in the Hamiltonian are again indicated. }
\label{groundstateAT_shade} 
\end{figure}

The ground state structure, and the mechanism of anisotropic couplings which allows the plane spin flips,
is thus identical  to that in $H_{dual1}$, whose possible ground states on a single cube  we recall for comparison
in Figure~\ref{groundstateAT_shade}.
Flipping the third spin $\mu$ in $H_{dual2}$, which is absent in $H_{dual1}$, is replaced by flipping both the $\sigma$ and $\tau$ spins in $H_{dual1}$, consistent with the gauge transformation relating the two Hamiltonians.
In summary, the ground state structure of $H_{dual2}$ shows an interesting interplay between the gauge symmetry of Equation~(\ref{local_gauge}) and the subsystem symmetry. The  local gauge symmetry allows one to reduce the effective number of degrees of freedom and 
recover the ground state structure of $H_{dual1}$.

$H_{dual3}$ is a similar case since the geometrical arrangement of the couplings is similar in spite of the presence of the additional spins $U$ on the links.  Each of the $\sigma, \tau, \mu$ spins couple in two directions, which define the lattice planes in which they can be flipped without affecting the energy.

\section{Indicative Monte-Carlo}
 
\begin{figure}[h!]
\centering
\includegraphics[height=8cm]{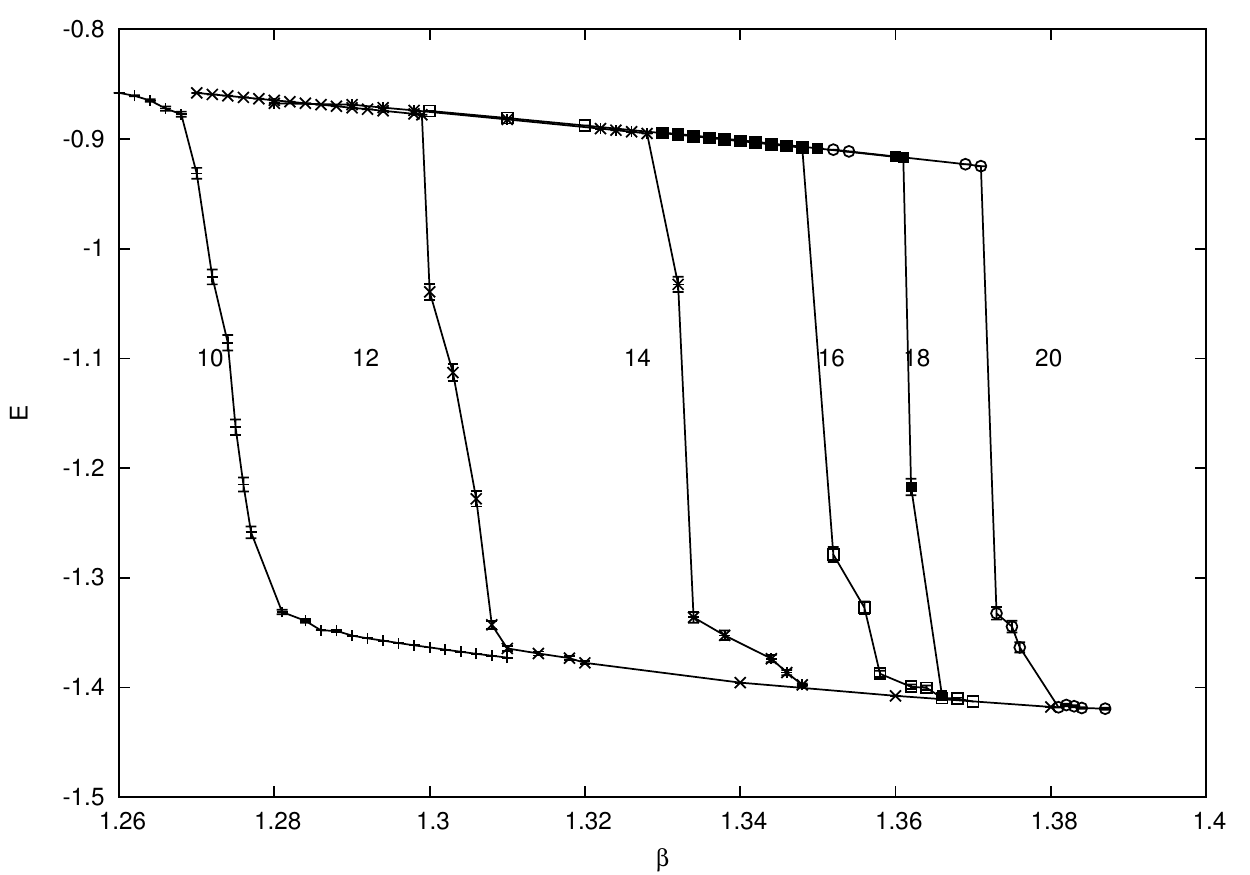}
\caption{The energy for $H_{dual2}$ on lattices ranging from $10^3$ to $20^3$ from left to right. The lines joining the data points are drawn to guide the eye. Data from $H_{dual1}$ is essentially identical.}
\label{E02} 
\end{figure}

Low precision Monte-Carlo simulations using simple Metropolis updates found  a first order phase transition in both $H_{dual2}$ and $H_{dual1}$ 
in the region $\beta \simeq 1.3-1.4$ \cite{5,5b}. Much higher precision simulations using multicanonical methods were later   carried out for 
the original $3d$ plaquette Ising model $H_{\kappa=0}$ and the Ashkin--Teller dual $H_{dual1}$ in order to accurately determine the transition point ($\beta^{\infty} = 1.31328(12)$ in the case of the dual model) and confirm the non-standard finite size scaling that is a consequence of the exponential degeneracy of the low temperature phase \cite{1sta,1stb,1stc,1std,1ste}.
Even with the modest statistics and the use of a Metropolis update in the simulations  of  \cite{5,5b} a sharp drop in the energy, as would be expected for a first order transition,  is clearly visible 
in the region of the transition point. A plot of the energy is shown for various lattice sizes 
in Figure~\ref{E02} for $H_{dual2}$ and the values for $H_{dual1}$ are essentially identical. 
The first order nature of the transition for  $H_{dual2}$ and $H_{dual1}$  can be further confirmed by 
observing a dual peak structure in the energy histogram $P(E)$ near the transition point and a non-trivial value  of Binder's energy cumulant
\begin{equation}
	U_E = 1 - \frac{\langle E^4 \rangle}{3  \langle E^2 \rangle^2} 
\end{equation}
as a consequence of the shape of $P(E)$ \cite{1sta,5b}. 

Based on these observations, and allowing for a factor of $1/2$ in our definitions of $H_{dual1}$ and $H_{dual2}$ in \cite{5,5b}, we would expect to see
a transition in $H_{dual3}$ at the the value of $\tilde \beta$ 
found by inverting the decoration transformation, namely $\frac{1}{2}  \cosh^{-1}(\exp(1.3-1.4))=0.99 - 1.04$
in the thermodynamic limit. 
To confirm this expectation, we carried out Monte-Carlo simulations using $10^3, 12^3, 16^3$ and $18^3$  lattices with periodic boundary conditions for all spins 
 at various temperatures, again  with a simple Metropolis update. After  an appropriate number of thermalization sweeps, $10^7$ measurement sweeps were carried out at each lattice size for each temperature.  

\begin{figure}[h!]
\centering
\includegraphics[height=8cm]{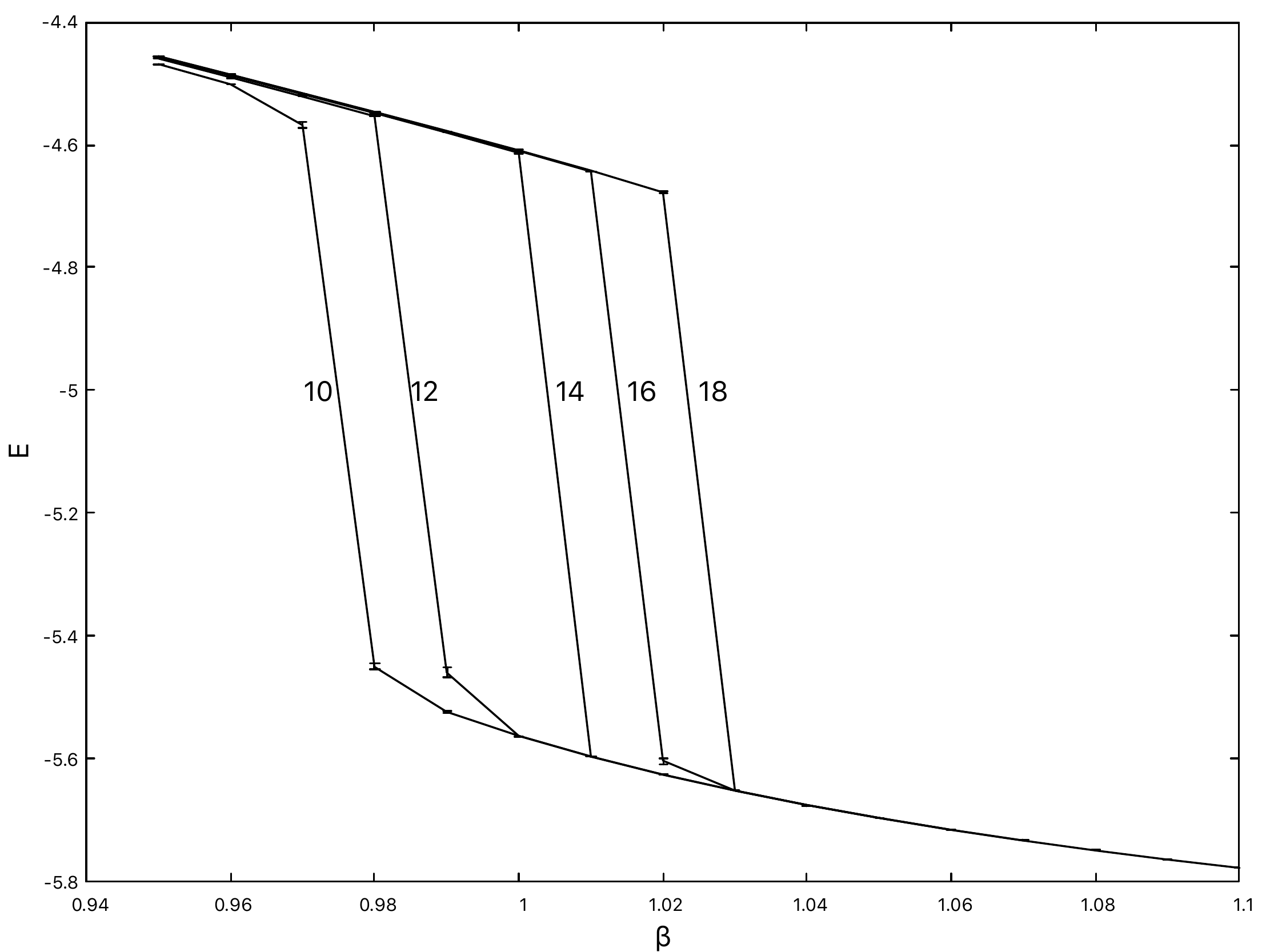}
\caption{The energy for $H_{dual3}$ on lattices ranging from $10^3$ to $18^3$ from left to right. The lines joining the data points are drawn to guide the eye.}
\label{E0c} 
\end{figure}

\begin{figure}[h!]
\centering
\includegraphics[height=7cm]{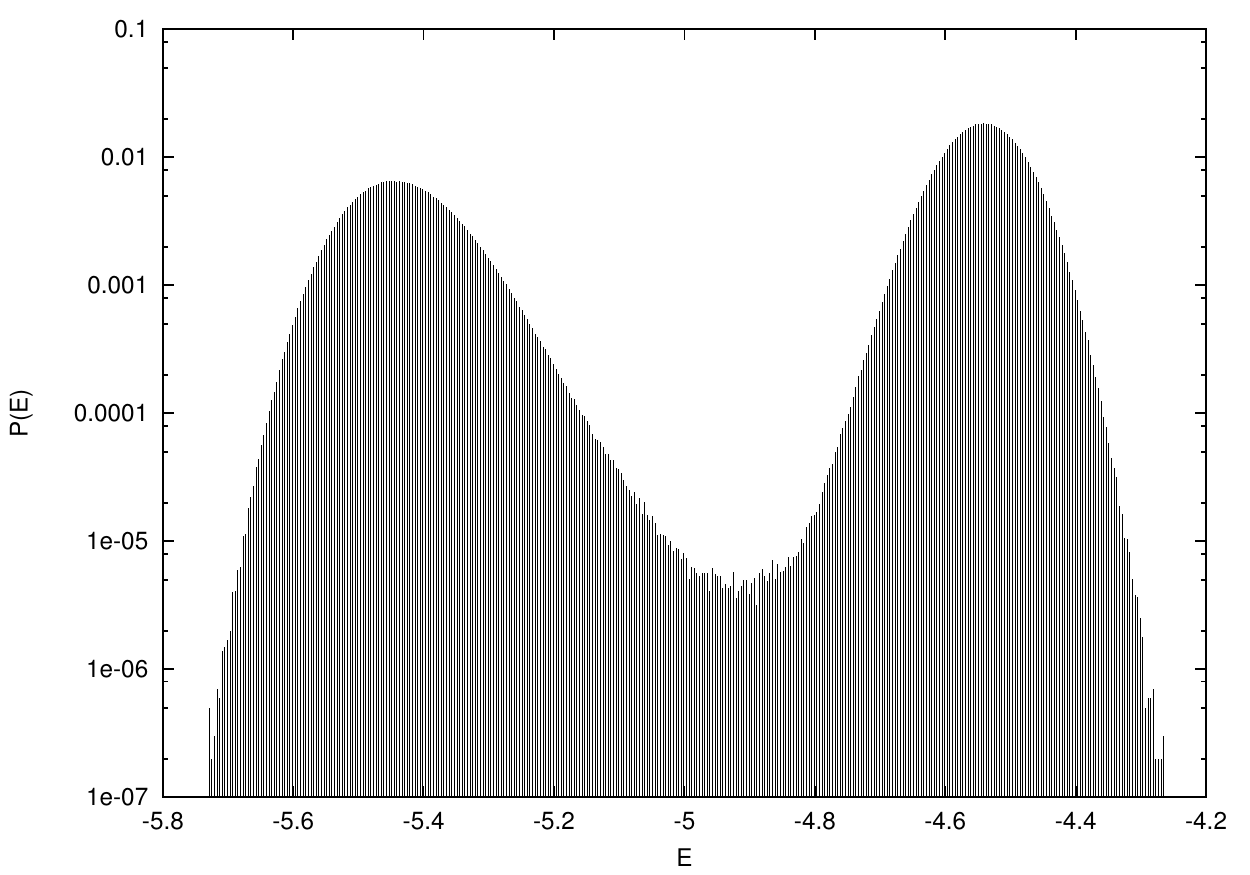}
\caption{{The energy histogram} 
 $P(E)$ for $H_{dual3}$ close to the estimated transition point at $\beta \simeq 0.97$ on a $10^3$ lattice.}
\label{PE} 
\end{figure}

Looking at measurements of the energy from our simulations of $H_{dual3}$ in Figure~\ref{E0c} we can see that a similar sharp drop in the energy consistent with a 
first order transition is still present.
The observed finite size estimates for the  transition temperatures $\beta_c(L)$ agree with those calculated by transforming
the values from Figure~\ref{E02} using the decoration relation, e.g., for $L=10$ we would expect  $\beta_c (10) \simeq \frac{1}{2}\cosh^{-1} [ \exp (1.27) ] \simeq 0.97$, as found directly in the simulation shown in Figure~\ref{E0c}.
Further evidence for a first order transition with $H_{dual3}$, as noted above for the other dual Hamiltonians, can be garnered by looking at the energy histogram $P(E)$ to discern a dual peak structure near the transition point. In Figure~\ref{PE}  $P(E)$ is shown close to the estimated transition point for $L=10$ at $\beta \simeq 0.97$ and there is clear evidence of two peaks.
The relatively low statistics and the use of a Metropolis update  for the data presented here for $H_{dual2}$ and $H_{dual3}$ mean that a high accuracy extrapolation using the correct $1/L^2$ finite size scaling for the transition point is not feasible, which would require more extensive multicanonical simulations along the lines of \cite{1sta,1stb,1ste}. 
Nonetheless, the agreement of the suitably transformed finite size lattice transition points  in the Monte-Carlo simulations confirm  that $H_{dual3}$ and $H_{dual2}$ are related by 
the decoration transformation and 
the first order nature of the transition for $H_{dual3}$ is clear  from the dual peak form of $P(E)$ near the transition point in Figure~\ref{PE}.

\section{Dynamics}

Another   interesting feature  of the original plaquette Hamiltonian $H_{\kappa=0}$ is its dynamical behaviour. It possesses a region of strong metastability around the first order phase transition
and displays glassy characteristics at lower temperatures \cite{4a,4b,4c,4d,4e,4f,4g} with non-trivial ageing properties.
We found that the Ashkin--Teller
dual Hamiltonian $H_{dual1}$ also shares these characteristics since it failed to relax to the equilibrium minimum energy of $E = -1.5$ when cooled quickly from a hot start \cite{5}. Note that in this case, since we are exploring the real time dynamics of the system, simulations with a Metropolis update are preferable to more sophisticated algorithms. 
 
\begin{figure}[h!]
\centering
\includegraphics[height=7cm]{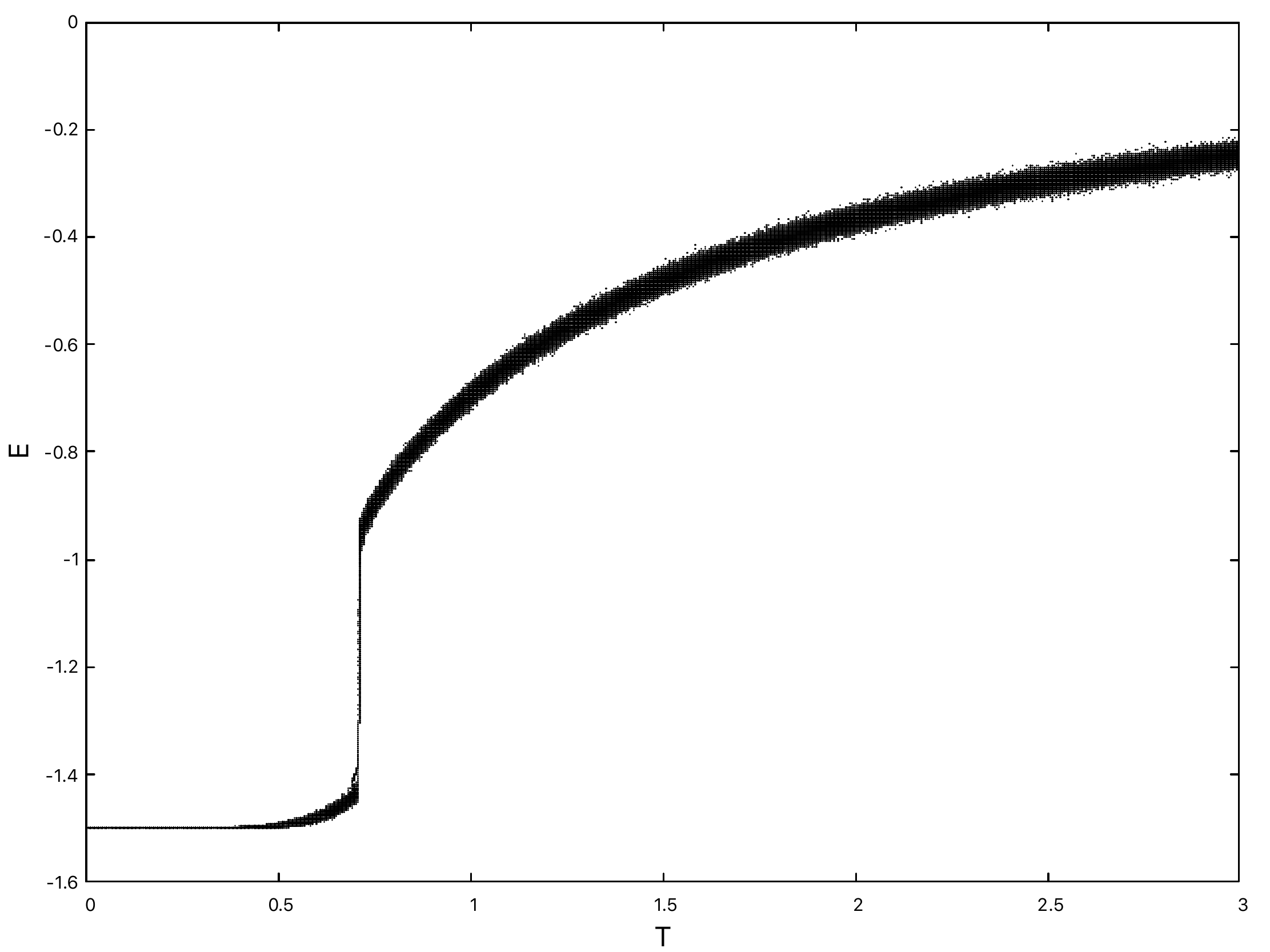}
\caption{The time series of energy measurements obtained from cooling $20^3, 60^3$ and $80^3$ lattices from a hot start at $T=3.0$
at a rate of $\delta T = 0.00001$ per sweep. The traces are effectively indistinguishable.}
\label{r00001} 
\end{figure}

\begin{figure}[h!]
\centering
\includegraphics[height=6cm]{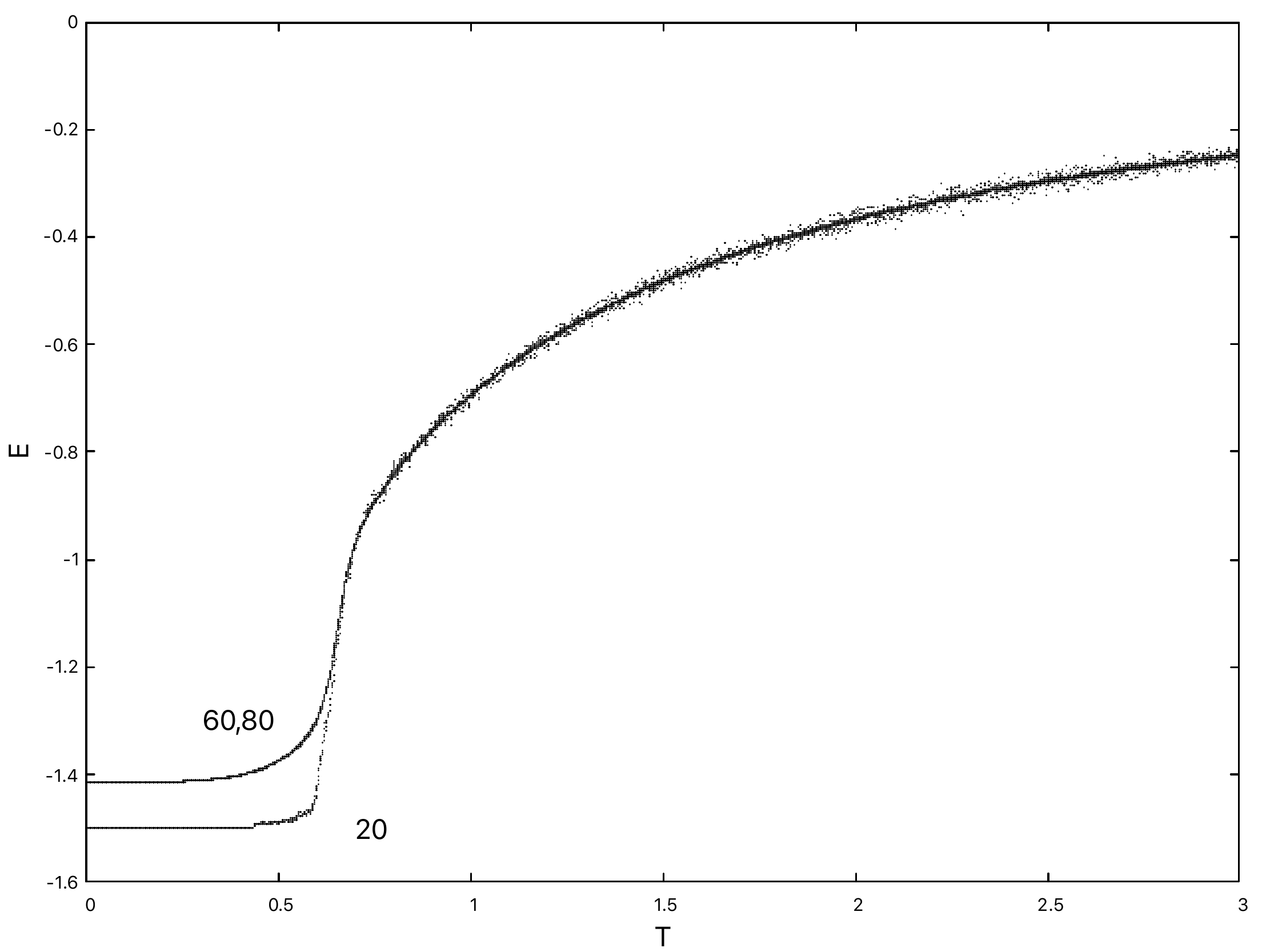}
\caption{The time series of energy measurements obtained from cooling $20^3, 60^3$ and $80^3$ lattices from a hot start
at a rate of $\delta T = 0.001$ per sweep.}
\label{r001} 
\end{figure}

$H_{dual2}$ displays identical behaviour under cooling to $H_{dual1}$. We consider
$20^3, 60^3$ and $80^3$ lattices which are first equilibrated in the high temperature phase at  $T=3.0$ and then cooled
at different rates to zero temperature. The energy time series is recorded during this process.
In Figure~\ref{r00001} we can see that with a slow cooling rate of $\delta T = 0.00001$ per sweep, the model still relaxes 
to a ground state with $E=-1.5$ for all the lattice sizes. 
However, as can be seen in Figure~\ref{r001} with a faster cooling rate of $\delta T = 0.001$ per sweep the model no longer relaxes to the ground state energy of $E=-1.5$, but is trapped at a higher value, which is around $-1.415$ for the larger two ($60^3$ and $80^3$) lattices.
Whether the observed behaviour under cooling is a sign of genuine glassiness in $H_{\kappa=0}$  or not remains a matter of debate
and similar considerations would apply to the dual Hamiltonians discussed here.

\section{Discussion}

Motivated by recent work on the quantum X-Cube Hamiltonian and related dual models \cite{odd} we revisit various  formulations 
of classical spin Hamiltonians dual to the  $3d$ plaquette Ising model.
We describe the following chain of relations between these    models
\bea
H_{dual3} &=& - \sum_{\langle ij \rangle} \left( \sigma_i U^{1}_{ij} \sigma_j  + \mu_i U^{1}_{ij} \mu_j \right) 
- \sum_{\langle ik \rangle } \left( \tau_i U^{2}_{ik} \tau_k +  \mu_i  U^{2}_{ik} \mu_k \right)  \nonumber \\
&-& \sum_{ \langle jk \rangle} \left( \sigma_j U^{3}_{jk} \sigma_k  + \tau_j U^{3}_{jk} \tau_k  \right) \nonumber \\
{} \nonumber \\
&{}&\longrightarrow \textrm{(Un)Decoration} \longrightarrow \nonumber \\
{} \nonumber \\
H_{dual2} &=& - \sum_{\langle ij \rangle} \sigma_i \sigma_j \mu_i \mu_j
- \sum_{\langle ik \rangle } \tau_i \tau_k  \mu_i \mu_k 
- \sum_{ \langle jk \rangle} \sigma_j \sigma_k \tau_j \tau_k \nonumber \\
{} \nonumber \\
&{}& \longrightarrow 
\textrm{Gauge-Fixing} \longrightarrow
 \\
{} \nonumber \\
H_{dual1} &=& - \sum_{ \langle ij \rangle} \sigma_{i}  \sigma_{j} 
- \sum_{ \langle ik \rangle } \tau_{i}  \tau_{k} 
-  \sum_{\langle jk \rangle} \sigma_{j} \sigma_{k} \tau_{j}  \tau_{k} \nonumber \\
{} \nonumber \\
&{}& \longrightarrow 
\textrm{Non-Ising variables} \longrightarrow \nonumber \\
{} \nonumber \\
H_{dual0} &=& - \sum_{\langle ij \rangle} \sigma_{i}  \sigma_{j} 
- \sum_{\langle ik \rangle } \tau_{i}  \tau_{k} 
-  \sum_{ \langle jk \rangle} \eta_{j} \eta_{k} \nonumber \\
{} \nonumber \\
&{}& \longrightarrow 
\textrm{Kramers--Wannier duality} \longrightarrow \nonumber \\
{} \nonumber \\
H_{\kappa=0} &=&  -  \sum_{\Box}\sigma_{i} \sigma_{j}\sigma_{k} \sigma_{l} \nonumber
\eea
where we have indicated the operations relating the various Hamiltonians.  A variant of the  decoration transformation in which edge spins are summed out relates $H_{dual3}$ to $H_{dual2}$. 
In transforming $H_{dual3}$  to $H_{dual2}$ the coupling is therefore transformed as $\beta =  \frac{1}{2} \ln [ \cosh ( 2 \tilde \beta )]$.
The gauge-invariant nature of $H_{dual3}$ due to the presence of both edge and vertex spins
leaves an echo in the vertex spin gauge symmetry of $H_{dual2}$, which in turn ensures the equivalence of $H_{dual2}$ and $H_{dual1}$ via
a gauge-fixing. Allowing non-Ising spins gives a final equivalence between the dual models
$H_{dual1}$ and $H_{dual0}$ and a standard Kramers--Wannier  duality transformation then takes us back to the  $3d$ plaquette Ising Hamiltonian of $H_{\kappa=0}$ where the story began.

The planar subsystem symmetry of this $3d$ plaquette Ising Hamiltonian $H_{\kappa=0}$ remains a feature of the various dual Hamiltonians and affects the finite size scaling properties at the first order transition displayed by  these models, just as with    $H_{\kappa=0}$.  The nature of the order parameter 
for the various duals, and indeed $H_{\kappa=0}$ itself, remains to be satisfactorily clarified. An attempt at this has been made for $H_{\kappa=0}$ in \cite{fuki-nuke} and indeed appeared to give sensible numerical results in \cite{1sta,1stb,1stc,1std,1ste}. This was based on the observation by Suzuki et al. \cite{suzuki1,suzuki2,suzuki3} that an anisotropic version of the $3d$ plaquette Ising model with open boundary conditions could be transformed to an uncoupled stack of $2d$ Ising models and suggested that a two spin correlator summed perpendicular to lattice planes might still serve as an order parameter in the isotropically coupled case of $H_{\kappa=0}$. It would be more satisfactory to have a less heuristic approach to an order parameter based on a clearer understanding of the nature of the low temperature order in 
the  $3d$ plaquette Ising Hamiltonian.
In this respect a study of the order parameters in the various dual models might be helpful.

In a similar vein, the principal interest in \cite{odd} was actually  investigating ``odd'' variants of fracton models, in which the signs of some of the terms in the Hamiltonians were reversed,  leading to frustrated models. The geometrical nature of the order in such frustrated spin models would be of interest in the classical case too.   Finally, the dynamics of the various classical Hamiltonians discussed here display glassy features. The question of whether the glassy dynamics of the quasiparticle excitations in the quantum models \cite{glassy-fracton1,glassy-fracton2} offers any insights into this behaviour may be worth pursuing.

\section{Acknowledgements}
The work of Ranasinghe  P.K.C.M. Ranasinghe was supported by a Commonwealth Academic Fellowship {\bf LKCF-2010-11}
and would like to thank Heriot-Watt University for hospitality when this work was initiated.



\end{document}